# Origin and control of high-temperature ferromagnetism in semiconductors

Shinji Kuroda[1*], Nozomi Nishizawa[1], Kôki Takita[1], Masanori Mitome[2], Yoshio Bando[2], Krzysztof Osuch[3,4] and Tomasz Dietl[5,6*]

[1]Institute of Materials Science, University of Tsukuba, 1-1-1 Tennoudai, Tsukuba, Ibaraki 305-8573, Japan

[2]Advanced Materials and Nanomaterials Laboratories, National Institute for Materials Science (NIMS), Namiki 1-1, Tsukuba, Ibaraki 305-0044, Japan

[3]Faculty of Physics, Warsaw University of Technology, Koszykowa 75, PL 00-662 Warszawa, Poland

[4]Department of Physics, University of South Africa, P. O. Box 392, Pretoria 0003, South Africa

[5]Laboratory for Cryogenic and Spintronic Research, Institute of Physics, Polish Academy of Sciences, and ERATO Semiconductor Spintronics JST Project, al. Lotników 32/46, PL 02-668, Warszawa, Poland

[6]Institute of Theoretical Physics, Warsaw University, PL 00-681 Warszawa, Poland

*e-mail: kuroda@ims.tsukuba.ac.jp; dietl@ifpan.edu.pl

**The extensive experimental and computational search for multifunctional materials has resulted in the development of semiconductor and oxide systems, such as (Ga,Mn)N, (Zn,Cr)Te, and $HfO_2$, which exhibit surprisingly stable ferromagnetic signatures despite having a small or nominally zero concentration of magnetic elements. Here, we show that the ferromagnetism of (Zn,Cr)Te, and the associated magnetooptical and magnetotransport functionalities, are dominated by the formation of Cr-rich (Zn,Cr)Te metallic nanocrystals embedded in the Cr-poor (Zn,Cr)Te matrix. Importantly, the formation of these nanocrystals can be controlled by manipulating the charge state of Cr ions during the epitaxy. The findings provide insight into the origin of ferromagnetism in a broad range of semiconductors and oxides, and indicate possible functionalities of these composite systems. Furthermore, they demonstrate a bottom-up method for self-organized nanostructure fabrication that is applicable to any system in which the charge state of a constituent depends on the Fermi-level position in the host semiconductor.**

Prompted by the demonstration of novel device principles[1] and theoretical suggestions[2], considerable research activities have been directed towards the development of functional ferromagnetic semiconductors. Comprehensive experimental efforts have been stimulated further by results of first-principles computations, which have substantiated the possibility of a robust ferromagnetism in a variety of systems[3,4], even in the absence of valence-band holes, which were originally regarded as being a necessary ingredient for high-temperature ferromagnetism in magnetically diluted semiconductors[2]. Indeed, a ferromagnetic response persisting up to above room temperature has been reported for a broad range of diluted magnetic semiconductors (DMS) and diluted magnetic oxides (DMO) containing minute amounts of magnetic ions[5-9] or even for materials built nominally of non-magnetic elements[10-12].

However, despite the apparent agreement between promising experimental findings and state-of-the-art computation expectations, the origin of abundant ferromagnetism in semiconductors and oxides has started to emerge as the most unsettled question in today's materials science and engineering[9]. Although a number of cases can be invoked here, we recall the most studied materials in this context: (Ga,Mn)N and (Zn,Co)O. In particular, experimental data for nominally intrinsic wurtzite $Ga_{1-x}Mn_xN$ reveal an astonishingly wide spectrum of properties: the material is either ferromagnetic with a reported Curie temperature, $T_C$, between 940 K (ref. 13) and 8 K (ref. 14) at $x = 0.06$ or its magnetization shows that the spin-spin coupling is dominated by antiferromagnetic interactions[15]. On the theoretical side, various implementations of density functional theory (DFT) predict[3,16,17] robust high-$T_C$ ferromagnetism in intrinsic $Ga_{1-x}Mn_xN$. However, when supplementing DFT with Monte Carlo simulations, the theoretical values of $T_C$ become much reduced, down to 55 K at $x = 0.05$ (ref. 18) or 30 K at $x = 0.06$ (ref. 19), with high $T_C$ being expected only in the presence of valence band holes[20].

In the case of (Zn,Co)O, the initial experimental observation[21] of the high-temperature ferromagnetism, supported by subsequent DFT results[3], stimulated a considerable experimental effort. It has been found that the ferromagnetism appears in the presence of deviations from stoichiometry, which produce donor defects[9,22]. This, together with more recent DFT work[23] calling into question the possibility of ferromagnetism in the absence of the valence-band holes, led to the conclusion that long-range magnetic order reflects the percolation of bound magnetic polarons produced around donor defects[9]. However, the typical magnetic polaron energies are rather low[24] and, accordingly, no quantitative arguments supporting the polaron scenario have so far been put forward.

The case of ferromagnetism in (Zn,Cr)Te is particularly intriguing. Here, a quantitative agreement between the measured[25] and computed $T_C$ is found[18,26]. Furthermore, the excellent agreement between direct magnetization measurements and magnetic circular dichroism has been taken as the decisive argument for the intrinsic nature of ferromagnetism in this DMS alloy[27]. This is in apparent contradiction with the case of (Zn,Cr)Se, for which Cr-rich nanoparticles were suggested to account for ferromagnetic signatures[28]. Recently, some of us have found a strong dependence of the ferromagnetic properties of (Zn,Cr)Te on co-doping with shallow acceptors[29] and donors[30]. Similarly to the earlier reports on (Zn,Mn)O:N and (Zn,Co)O (ref. 22) as well as on (Ga,Mn)N:Mg,Si (refs 31,32), the data[29,30] were interpreted in terms of the effect of co-doping on the spin-spin coupling mediated by itinerant carriers in the $d$-band. At the same time, one of us has suggested[33] that in the systems in question co-doping by shallow impurities, by changing the charge state of magnetic ions, affects the aggregation of magnetic constituents rather than the spin-spin interaction.

Here, we present the results of a thorough structural characterization combined with superconducting quantum interference device magnetometry for zinc-blende (Zn,Cr)Te films, in which the position of the Fermi level is engineered either by co-doping or by controlled deviations from the stoichiometry. Our findings reveal that the dramatic changes of the apparent $T_C$ with doping[29,30] reflect a strong influence of the Fermi-level position on the uniformity of the Cr distribution over the zinc-blende lattice. We link this effect to the predicted[33] dependence of the solubility limit on the charge state of the relevant magnetic ions. This interpretation is validated by the known positions of the Cr, N, and I levels in the bandgap of ZnTe (refs 34-36) as well as by our *ab initio* computation, which shows strong changes in the interaction energy of Cr pairs in ZnTe with the Cr charge state. Therefore, our results shed entirely new light on the origin of high-temperature ferromagnetism in DMS and DMO as well as suggesting its possible application in memory and photonics devices. Moreover, our findings demonstrate the viability of the self-organized growth mode[33], in which the control over the Fermi-level position during epitaxy steers the bottom-up self-assembling of magnetic nanostructures in a semiconductor or oxide matrix.

**Determination of Cr distribution**

We prepared a series of $Zn_{1-x}Cr_xTe$ films grown by molecular beam epitaxy (MBE), in which either deviation from stoichiometry achieved by varying the Zn/Te flux ratios or co-doping by N acceptors and I donors control the Fermi-level position. To suppress the self-compensation by native defects, the growth of I-doped (N-doped) films was carried out under Zn-rich (Te-rich) conditions. We assessed the crystal structure and the distribution of constituent elements by transmission electron microscopy (TEM) and energy-dispersive X-ray spectroscopy (EDS), respectively. Figure 1 shows a typical example of cross-sectional TEM and electron diffraction images of the I-doped films. The lattice image exhibits mostly zinc-blende structure, but in some regions stacking faults along the {111} plane appear. As shown in Fig. 1b, c, the electron diffraction images of $Zn_{1-x}Cr_xTe$ near the interface of the buffer and near the surface are different. The diffraction close to the interface (Fig. 1b) exhibits the fundamental hexagonal spots of the zinc-blende structure and additional weak spots appear at one-third positions between the fundamental spots, corresponding to a triplet periodicity in the stacking sequence along the {111} plane of the observed stacking faults. In the diffraction close to the surface (Fig. 1c), zinc-blende hexagonal patterns with different orientations are superimposed, reflecting the deterioration of the crystal structure into a multidomain structure consisting of zinc-blende nanocrystals as the growth of the $Zn_{1-x}Cr_xTe$ layer proceeds. However, any apparent precipitates of other crystal structures have not been detected in either the lattice image or in electron diffraction.

Remarkably, spatially resolved EDS reveals that the co-doping and deviations from stoichiometry affect the spatial distribution of Cr over the zinc-blende lattice in a striking way. Figure 2 shows the mapping of the Cr $K_\alpha$ emission intensity of I-doped, N-doped, and undoped films with an average Cr concentration $x \approx 0.05$. The Cr distribution in the N-doped film (Fig. 2f) and in the undoped film grown under Te-rich conditions (Fig. 2e) is apparently homogeneous, whereas the Cr mapping images of the I-doped films (Fig. 2a-c) and the undoped film grown under Zn-rich conditions (Fig. 2d) exhibit non-uniform Cr distributions. As the characteristic length scale of the revealed mesoscopic fluctuations in the Cr concentration is smaller than the thickness of the probed sample, $t \approx 100$ nm, we can only determine the lower limit of the Cr concentration in the Cr-rich regions, $x > 0.1$, and the upper limit of the Cr-rich regions diameter, $d < 50$ nm (Fig. 2b,d) and $d < 30$ nm (Fig. 2a,c). We have actually succeeded in the preparation of a much thinner sample with the same I concentration as that in Fig. 2b, whose EDS image allows us to evaluate $d \leq 20$ nm and $x \geq 0.4$. The combined analysis of EDS mapping and TEM images confirms the zinc-blende crystal structure of the Cr-rich regions as well as the lack of correlation between the Cr distribution and the stacking faults visible in the lattice images. We conclude from these findings that the control over the growth conditions and/or the co-doping makes it possible to engineer nanocrystals of Cr-rich (Zn,Cr)Te embedded coherently in a Cr-poor (Zn,Cr)Te matrix.

We will now demonstrate that the Cr-rich (Zn,Cr)Te nanocrystals account for the previously reported ferromagnetic properties of (Zn,Cr)Te (refs 29,30). We have found that temperature and magnetic-field dependencies of magnetization, $M(T,H)$, and susceptibility, $\chi(T,H)$, exhibit features typical for superparamagnetic materials. Accordingly, we describe the system in terms of three characteristic temperatures: (1) the apparent Curie temperature, $T_C^{(app)}$, deduced from the Arrott plot analysis of the $M$-$H$ curves, (2) the paramagnetic Curie-Weiss temperature, $\Theta_p$, obtained from the $\chi^{-1}$-$T$ plots at $T > T_C^{(app)}$, and (3) the mean blocking temperature, $T_B$, determined from the cusp positions in the $M$-$T$ dependencies for zero-field-cooled magnetization measurements. By comparing the EDS data (Fig. 2) with the results of magnetic studies for the same series of samples (Fig. 3), we see that $T_C^{(app)}$ is low or even vanishes in the case of the films with a homogeneous Cr distribution (sample e: Te-rich and sample f: N-doped). On the other hand, a non-uniform Cr distribution (sample d: Zn-rich and samples a-c: I-doped) gives rise to higher values of $T_C^{(app)}$ and $\Theta_p$, both reaching 300 K for the films containing the largest Cr-rich nanocrystals (samples b and d). These high values of $T_C^{(app)}$ and $\Theta_p$ are consistent with the experimental results for zinc-blende CrTe epilayers showing $T_C$ = 327 K (ref. 37), a value setting an upper limit for the $T_C^{(app)}$ of the Cr-rich clusters. When the volume, $V$, of ferromagnetic nanocrystals decreases, $T_C^{(app)}$ diminishes, as its magnitude is determined by the upper limit of the blocking temperature, $T_B$, which is proportional to $V$. Similarly, $\Theta_p$ decreases, as the ratio of dangling magnetic bonds at the nanocrystal surface to the total number of the bonds increases with decreasing $V$. As the latter scales with $V^{1/3}$, $\Theta_p$ is expected to diminish slower with $V$ than $T_C^{(app)}$, in agreement with the results shown in Fig. 3. As both $T_C^{(app)}$ and $\Theta_p$ tend to vanish when the distribution of the Cr atoms becomes uniform, we assert that the high-temperature ferromagnetism of (Zn,Cr)Te appears if the growth conditions and/or co-doping promote the assembly of Cr-rich nanocrystals. This conclusion implies that the state-of-the art computations of $T_C$ in DMS[18,26], which show the agreement between $T_C^{(app)}$ and the theoretical $T_C$ obtained for the uniform Cr distribution in (Zn,Cr)Te, overestimate the magnitude of ferromagnetic couplings.

It is interesting to compare the diameter of the Cr-rich nanocrystals as revealed by EDS with that implied by the superparamagnetic characteristics[38], $d \approx (150\, k_B T_B/\pi K)^{1/3}$, where $k_B$ is the Boltzmann constant and the density of anisotropy energy $K = M_S H_A/2$. Assuming the Cr magnetic moment $\mu = 2.2\mu_B$ (ref. 25) and the Cr concentration within the nanocrystals $x = 1$, we obtain a saturation value of magnetization $M_S = 7.2 \times 10^2$ e.m.u. As nanocrystals should be single domain we can identify the anisotropy field, $H_A$, with the low-temperature coercivity force, typically of the order of 500 Oe (refs 25,29,30), which for $40 < T_B < 300$ K leads to $11 < d < 22$ nm. This finding is consistent with the EDS data, $d \leq 20$ nm, as quoted above, as well as with the notion that $T_C^{(app)}$ corresponds to the upper bound of $d$ and, hence, of the $T_B$ distribution. This distribution, together with possible interactions between neighbour nanocrystals, leads to a strong non-squareness of the hysteresis loops, behaviour specific to virtually all high-temperature DMS and DMO. The ratio of remanent magnetization, $M_r = M(H = 0)$, to $M_S$ immediately below $T_C^{(app)}$ suggests that the distribution tail containing about 10% of Cr nanocrystals contributes to $T_C^{(app)}$.

**Origin and control of Cr distribution**

To find out why the Cr distribution and the resulting magnetic properties depend on the growth conditions and co-doping, we note that except for Mn in II-VI compounds, the solubility of magnetic ions is rather low in semiconductors. Accordingly, if the concentration of magnetic ions exceeds the solubility limit, spinodal decomposition into regions with low and high density of magnetic ions – a process driven by attractive forces between magnetic ions in semiconductors – is expected[33,39,40]. The key observation here is that in most DMS and DMO, in contrast to non-magnetic alloys such as (Ga,In)N, the charge state of the constituents and, hence, the corresponding interaction energy, depends on the position of the Fermi level with respect to the band edges. Accordingly, the spinodal decomposition can be controlled by the growth conditions and co-doping[33,40]. In the case of Cr in ZnTe, according to magnetic resonance and optical studies[34,35], the $Cr^{2+}/Cr^{3+}$ donor state is located above the N-acceptor level[36], whereas the $Cr^{2+}/Cr^{1+}$ acceptor state resides about 1 eV higher but still below the I-donor level, as shown in Fig. 4a. This level diagram means that Cr is in the isoelectronic to Zn (neutral) 2+ charge state in intrinsic ZnTe. However, in the presence of shallow acceptors (Zn vacancies or N impurities) Cr assumes the 3+ charge state, whereas under doping by shallow donors (I-impurities), the charge state changes to +1. The Coulomb repulsion between the charged Cr ions should suppress the spinodal decomposition.

We quantify the model evaluating the total energy change, $E_d$, associated with bringing two Cr atoms to the nearest-neighbor positions in the cation sublattice in the zinc-blende crystal by *ab initio* all-electron DFT calculations, whose implementation was described in details elsewhere[41]. As shown in Fig. 4b, we find $E_d = -141$ meV in the case of the pair of neutral Cr atoms in ZnTe. This value can be compared with $E_d = + 21$ meV, which we obtain for a Mn pair, the results explaining a remarkably different solubility of Cr and Mn in ZnTe. However, the charging of Cr atoms by doping with shallow impurities generates Coulomb repulsion between Cr ions, which increases $E_d$ as shown in Fig. 4b. We argue, therefore, that this additional repulsion energy together with the entropy term overcompensate the gain in energy associated with the bonding of two transition-metal impurities. This impedes the spinodal decomposition and stabilises the epitaxy of a uniform alloy even if the fractional concentration is within the solubility gap for the isoelectronic constituents.

To link the experimental findings of Figs 2 and 3 with the theoretical results in Fig. 4, we recall that owing to the formation of Zn vacancies, the as-grown ZnTe is p-type, a typical concentration of native acceptors being $\sim 10^{18}$ cm$^{-3}$ (ref. 36). Accordingly, the growth under Zn-rich conditions or co-doping with an optimized I-donor concentration turns all Cr atoms into the 2+ neutral charge state, for which the Fermi level resides in the mid-gap region. Indeed, the material we fabricate under these conditions is semi-insulating. Neither double exchange nor band-carrier-mediated spin-spin interactions can explain the presence of high-temperature ferromagnetism of (Zn,Cr)Te in such a case. In contrast, the high $T_C$ value, specific to metallic CrTe, is readily explained by delocalization of the *d*-electrons inside the (Zn,Cr)Te nanocrystals with a large Cr density. The formation of such Cr-rich clusters is observed experimentally (Fig. 2) and expected theoretically for Cr in the neutral +2 charge state (Fig. 4). At the same time, in agreement with the model, the nanocrystal size and $T_C^{(app)}$ decrease, when $Cr^{+1}$ ions appear owing to co-doping with I donors in excess of $\sim 10^{18}$ cm$^{-3}$, the effect being visible in Figs 2 and 3. Similarly, $T_C^{(app)}$ diminishes under the presence of N acceptors or Zn vacancies, which generate Cr in the +3 charge state. In fact, in accordance with the present scenario, $T_C^{(app)}$ vanishes when the concentrations of Cr and N become comparable (Fig. 3 and ref. 29), so that all of the Cr atoms become charged and the spinodal decomposition stops completely. The decrease in $T_C^{(app)}$ begins when the net acceptor concentration is much smaller than the Cr density, which indicates that a single charge per nanocrystal can already prevent its further growth.

**Discussion and outlook**

In view of the rather limited solubility of magnetic impurities in semiconductors, we expect that the spinodal decomposition into regions with low and high concentrations of the magnetic constituent, revealed here for (Zn,Cr)Te, is a generic property of DMS and DMO. Indeed, such a chemical phase separation has also been found in annealed (Ga,Mn)As (ref. 42) as well as in as-grown epitaxial films of (Ga,Mn)N (ref. 43), (Al,Cr)N (ref. 44) and (Ge,Mn) (refs 45,46). Interestingly, the nanocrystals fabricated in this way exhibit novel ferromagnetic characteristics, not encountered for the corresponding materials in bulk forms, as the nanocrystal composition and structure are imposed by the relevant host. It is expected that the identification of the actual chemical nature of nanocrystals in particular matrices, as well as the determination of the mechanisms accounting for the necessarily high blocking temperatures making the ferromagnetic-like signatures survive up to high temperatures, will attract considerable attention in future.

We note that uncompensated spins in antiferromagnetic nanoparticles can also produce sizable spontaneous magnetization at high temperatures[47]. The Fermi-level-dependent assembly of antiferromagnetic CoO-rich coherent nanocrystals may therefore account for the ferromagnetic-like behaviour of (Zn,Co)O as well as explain why the spontaneous magnetic moment per Co is so small[9,21,22]. It has also been argued that contamination by magnetic nanoparticles accounts for ferromagnetic signatures of carbon[48] and $HfO_2$ (ref. 49). Further progress in spatially resolved chemical analysis as well as in the simultaneous modelling of $T_C$ and electrical conductance will show whether the above room-temperature ferromagnetism is possible in materials containing no magnetic elements.

Remarkably, our anomalous Hall effect and magnetic circular dichroism studies on (Zn,Cr)Te, summarized in Supplementary Information, Figs S3 and S4, confirm that the presence of magnetically active metallic nanocrystals leads to enhanced magnetotransport[38] and magnetooptical[42] properties over a wide spectral range. This opens the way to various applications of such hybrid semiconductor/ferromagnetic systems[33], particularly for volumetric magnetic and holographic recording, optical isolation and modulation, magnetic-field sensing, and spin injection. In this context, the development of methods affecting nanocrystal characteristics and distribution is timely. Self-organized nano-assembly controlled by the Fermi-level position, demonstrated here for (Zn,Cr)Te, offers a method that is applicable to a broad variety of compounds, in which growth conditions determine the charge state of a constituent.

It is instructive to compare the discussed wide-bandgap DMS and DMO, in which magnetic impurities give rise to deep levels, with the case of Mn-doped GaAs, InAs, GaSb and InSb. In those systems, owing to the relatively shallow character of Mn acceptors and the large Bohr radius, the holes reside in the valence band[2]. Thus, the Mn atoms are charged, which – according to our model – reduces their aggregation and makes it possible to deposit, by low-temperature epitaxy, a uniform alloy with a composition beyond the solubility limit. Co-doping with shallow donors, by reducing the free-carrier screening, will enhance the repulsions among Mn, and allow us to fabricate homogenous layers with even greater $x$. On the other hand, co-doping by shallow acceptors, along with the donor formation by a self-compensation mechanism[8], will enforce the screening and, hence, allow the nanocrystal formation. Importantly, the effect of self-compensation, which consists of the appearance of compensating point defects once the Fermi level reaches an appropriately high position in the conduction band (donor doping) or low energy in the valence band (acceptor doping), is not relevant in wide-bandgap DMS and DMO, where the Fermi level is pinned by magnetic impurities in the gap region.

In summary, we argue that the puzzling high-temperature ferromagnetic-like response detected in a number of DMS and DMO containing no hole carriers[5-9] originates from nanocrystals with a large density of magnetic ions. This is in accord with the results of recent materials characterization and computational design studies that also accentuate the key importance of itinerant valence-band holes in carrying a strong long-range ferromagnetic coupling between randomly distributed localized spins in spatially uniform DMS and DMO, such as (Ga,Mn)As. According to our findings, the formation of the magnetic nanocrystals can be manipulated by the growth conditions and co-doping. We expect that this bottom-up nanofabrication method will achieve a similar significance in DMS and DMO research and applications as the exploitation of strain fields has accomplished in the self-organized growth of semiconductor quantum dots.

**METHODS**

**MBE growth and characterization**

Epitaxial layers of $Zn_{1-x}Cr_x Te$ were grown by conventional MBE using solid sources of Zn, Cr and Te. A ZnTe buffer layer (thickness ~ 700 nm) was first grown on a GaAs (001) substrate to relax the large lattice mismatch, and then a $Zn_{1-x}Cr_x Te$ layer (~ 300 nm) was deposited. Iodine (I) was used as an n-type charged impurity using a solid source of $CdI_2$, and nitrogen (N) was used as a p-type charged impurity using a $N_2$ gas source excited by radiofrequency plasma. I-doped (N-doped) $Zn_{1-x}Cr_x Te$ epilayers were grown in Zn-rich (Te-rich) conditions, to suppress the formation of native defects due to the effect of self-compensation. Undoped $Zn_{1-x}Cr_x Te$ epilayers were grown either in the Te-rich or Zn-rich condition. The substrate temperature was kept at 300°C during growth.

The Cr composition, *x*, was estimated employing an electron probe microanalyser (EPMA) with a low acceleration voltage to probe only the $Zn_{1-x}Cr_x Te$ layer. The iodine concentration, [I], was measured using secondary-ion mass spectroscopy (SIMS) with an ion-implanted ZnTe wafer as a reference.

**TEM and EDS analysis**

High-resolution TEM analysis was carried out in an energy-filtering analytical TEM with an accelerating voltage of 300 keV (JEM-3100FEF, JEOL) (ref. 50). A cross-sectional piece with a thickness $t \sim 100$ nm was cut from the grown film using focused ion beam. The composition of the constituent elements was investigated using EDS. The Cr distribution was surveyed by mapping the intensity of the Cr $K_\alpha$ emission in the scanning TEM mode. The resultant spatial resolution in the mapping plane was about 5 nm. In samples with non-uniform Cr distributions (Fig. 2a-d), the typical size of Cr-rich regions was given by 30-50 nm (Fig. 2b,d) or 10-30 nm (Fig. 2a,c), and the Cr composition in a Cr-rich region was estimated to be around $x \sim 0.1$ from the EDS spot analysis. However, these values should be considered as the upper or lower limit of the actual sizes and Cr compositions, respectively, as the measured values have been averaged over the total thickness of the probed pieces. For one of the I-doped samples (sample b in Fig. 3), we carried out the same EDS analysis on a thinner piece ($t \sim 50$ nm) prepared by Ar ion milling. As a result, smaller sizes (10-20 nm) and a higher Cr composition ($x \sim 0.4$) of Cr-rich regions were observed in this sample (see Supplementary Information, Fig. S1).

**Magnetization measurement**

The magnetization values were obtained by means of a superconducting quantum interference device magnetometer (MPMS-5SPL, Quantum Design). The measurements were carried out in the reciprocating sample measurement mode with the magnetic field applied perpendicular to the film plane. The contribution from the GaAs substrate was subtracted. It was verified that magnetization measurements under the magnetic field parallel to the plane provided similar results. From the temperature and field dependencies of magnetization, $M(T,H)$, three characteristic temperatures describing magnetic properties were deduced: the apparent Curie temperature, $T_C^{(app)}$, was determined from the Arrott plot analysis of the *M-H* curves; the paramagnetic Curie-Weiss temperature, $\Theta_p$, was deduced from a linear extrapolation of the $\chi^{-1}$-*T* plot; and the blocking temperature, $T_B$, was determined from the position of the maximum in the *M-T* dependence obtained after the zero-field-cooled process (see Supplementary Information, Fig. S2).

**References**

1. Ohno, Y., Young, D. K., Beschoten, B., Matsukura, F., Ohno, H. & Awschalom, D. D. Electrical spin injection in a ferromagnetic semiconductor heterostructure. *Nature* **402**, 790-792 (1999).
2. Dietl, T., Ohno, H., Matsukura, F., Cibert, J. & Ferrand, D. Zener model description of ferromagnetism in zinc-blende magnetic semiconductors. *Science* **287**, 1019-1022 (2000).
3. Sato K. & Katayama-Yoshida, H. First principles materials design for semiconductor spintronics. *Semicond. Sci. Technol.* **17** 367–376 (2002), and references therein.
4. Sandratskii, L. M. & Bruno, P. Electronic structure, exchange interactions, and Curie temperature in diluted III-V magnetic semiconductors: (GaCr)As, (GaMn)As, (GaFe)As. *Phys. Rev. B* **67**, 214402 (2003).
5. Pearton, S. J. *et al*. Wide band gap ferromagnetic semiconductors and oxides. *J. Appl. Phys.* **93**, 1-13 (2003).
6. Fukumura, T., Toyosaki, H. & Yamada, Y. Magnetic oxide semiconductors, *Semicond. Sci. Technol.* **20**, S103-S111 (2005).
7. Liu, C., Yun, F. & Morkoç, H. Ferromagnetism of ZnO and GaN: a review. *J. Mater. Sci.: Materials in Electronics* **16**, 555-597 (2005).
8. MacDonald, A. H., Schiffer, P. & Samarth N. Ferromagnetic semiconductors: moving beyond (Ga,Mn)As. *Nature Mater.* **4**, 195-202 (2005).
9. Chambers, S. A. *et al.* Ferromagnetism in oxide semiconductors. *Materials Today* **9**, 28-35 (2006), and references therein.
10. Young, D. P. *et al.* High-temperature weak ferromagnetism in a low-density free-electron gas. *Nature* **397**, 412-414 (1999).
11. Makarova, T. L. *et al.* Magnetic carbon. *Nature* **413**, 716-718 (2001).
12. Venkatesan, M., Fitzgerald, C. B. & Coey, J. M. D. Unexpected magnetism in a dielectric oxide. *Nature* **430**, 630 (2004).
13. Sonoda, S., Shimizu, S., Sasaki, T., Yamamoto, Y. & Hori H. Molecular beam epitaxy of wurtzite (Ga,Mn)N films on sapphire (0001) showing the ferromagnetic behaviour at room temperature. *J. Cryst. Growth* **237-239**, 1358-1362 (2002).
14. Sarigiannidou, E. *et al.* Intrinsic ferromagnetism in wurtzite (Ga,Mn)N semiconductor. *Phys. Rev. B* **74**, 041306(R) (2006).
15. Zając, M., Gosk, J., Kamińska, M., Twardowski, A., Szyszko, T. and & Podsiadło, S. Paramagnetism and antiferromagnetic $d$–$d$ coupling in GaMnN magnetic semiconductor. *Appl. Phys. Lett.* **79**, 2432-2434 (2001).
16. Kronik, L., Jain, M. & Chelikowsky, J. R. Electronic structure and spin polarization of $Mn_xGa_{1-x}N$. *Phys. Rev. B* **66**, 041203 (R) (2002).

17. Mahadevan, P. & Zunger, A. Trends in ferromagnetism, hole localization, and acceptor level depth for Mn substitution in GaN, GaP, GaAs, and GaSb . *Appl. Phys. Lett.* **85**, 2860-2862 (2004).
18. Bergqvist, L. *et al.* Magnetic percolation in diluted magnetic semiconductors. *Phys. Rev. Lett.* **93**, 137202 (2004).
19. Sato, K., Schweika, W., Dederichs, P. H. & Katayama-Yoshida, H. Low-temperature ferromagnetism in (Ga,Mn)N: *ab initio* calculations. *Phys. Rev. B* **70**, 201202(R) (2004).
20. Schulthess, T. C., Temmerman, W. M., Szotek, Z., Butler, W. H. & Stocks, G. M. Electronic structure and exchange coupling of Mn impurities in III-V semiconductors. *Nature Mater.* **4**, 838- 844 (2005).
21. Ueda, K., Tabata, H. & Kawai, T. Magnetic and electric properties of transition-metal-doped ZnO films. *Appl. Phys. Lett.* **79**, 988-990 (2001).
22. Kittilstved, K. R., Norberg, N. S. & Gamelin, D. R. Chemical manipulation of high-$T_C$ ferromagnetism in ZnO diluted magnetic semiconductors. *Phys. Rev. Lett.* **94**, 147209 (2005).
23. Spaldin, N. A., Search for ferromagnetism in transition-metal-doped piezoelectric ZnO. *Phys. Rev. B* **69**, 125201 (2004).
24. Dietl, T. & Spałek, J. Effect of fluctuations of magnetization on the bound magnetic polaron: comparison with experiment. *Phys. Rev. Lett.* **48**, 355-358 (1982).
25. Saito, H., Zayets, V., Yamagata, S. & Ando, K. Room-temperature ferromagnetism in a II-VI diluted magnetic semiconductor $Zn_{1-x}Cr_xTe$. *Phys. Rev. Lett.* **90**, 207202 (2003).
26. Fukushima, T., Sato, K., Katayama-Yoshida, H. & Dederichs, P. H. Theoretical prediction of Curie temperature in (Zn,Cr)S, (Zn,Cr)Se and (Zn,Cr)Te by first principles calculations. *Jpn. J. Appl. Phys.* **43**, L1416–L1418 (2004).
27. Ando, K. Seeking room-temperature ferromagnetic semiconductors, *Science* **312**, 1883-1885 (2006).
28. Karczewski, G. *et al*. Ferromagnetism in (Zn,Cr)Se layers grown by molecular beam epitaxy. *J. Supercond./Novel Magnetism* **16**, 55-58 (2003).
29. Ozaki, N. *et al.* Suppression of ferromagnetism due to hole doping in $Zn_{1-x}Cr_xTe$ grown by molecular beam epitaxy. *Appl. Phys. Lett.* **87**, 192116 (2005).
30. Ozaki, N. *et al.* Significant enhancement of ferromagnetism in $Zn_{1-x}Cr_xTe$ doped with iodine as an n-type dopant. *Phys. Rev. Lett.* **97**, 037201 (2006).
31. Reed, M. J. *et al.* Effect of doping on the magnetic properties of GaMnN: Fermi level engineering. *Appl. Phys. Lett.* **86**, 102504 (2005).
32. Kane, M. H. *et al.* Correlation of the structural and ferromagnetic properties of $Ga_{1-x}Mn_xN$ grown by metalorganic chemical vapor deposition. *J. Crystal Growth* **287**, 591–595 (2006).

33. Dietl, T. Self-organized growth controlled by charge states of magnetic impurities. *Nature Mater.* **5**, 673 (2006).
34. Godlewski, M. & Kamińska, M. The chromium impurity photogeneration transitions in ZnS, ZnSe and ZnTe. *J. Phys. C: Solid State Phys.* **13**, 6537-6545 (1980).
35. Dziesiaty, J. *et al.* The chromium impurity in ZnTe: changes of the charge state detected by optical and EPR spectroscopy. *Z. Phys. Chem.* **201**, S63-S73 (1997).
36. Baron, T., Saminadayar, K. & Magnea, N. Nitrogen doping of Te-based II-VI compounds during growth by molecular beam epitaxy. *J. Appl. Phys.* **83**, 1354-1370 (1998).
37. Sreenivasan, M. G. *et al.* Zinc-blende structure of CrTe epilayers grown on GaAs. *IEEE Trans. Magn.* **42**, 2691-2693 (2006).
38. Shinde, S. R. *et al.* Co-occurrence of superparamagnetism and anomalous Hall effect in highly reduced cobalt-doped rutile $TiO_{2-\delta}$ films. *Phys. Rev. Lett.* **92**, 166601 (2004).
39. Fukushima, T., Sato, K., Katayama-Yoshida, H. & Dederichs, P. H. *Ab initio* study of spinodal decomposition in (Zn,Cr)Te. *Phys. Status Solidi (a)* **203**, 2751-2755 (2006).
40. Ye, L.-H. & Freeman, A. J. Defect compensation, clustering, and magnetism in Cr-doped anatase $TiO_2$. *Phys. Rev. B* **73**, 081304(R) (2006).
41. Osuch, K., Lombardi, E. B. & Adamowicz, L. Palladium in GaN: a 4d metal ordering ferromagnetically in a semiconductor. *Phys. Rev. B* **71**, 165213 (2005).
42. Yokoyama, M., Yamaguchi, H., Ogawa, T. & Tanaka, M. Zinc-blende-type MnAs nanoclusters embedded in GaAs. *J. Appl. Phys.* **97**, 10D317 (2005).
43. Martinez-Criado, G. *et al.* Mn-rich clusters in GaN: hexagonal or cubic symmetry? *Appl. Phys. Lett.* **86**, 131927 (2005).
44. Gu, L. *et al.* Characterization of Al(Cr)N and Ga(Cr)N diluted magnetic semiconductors. *J. Magn. Magn. Mater.* **290-291**, 1395-1397 (2005).
45. Jamet, M. *et al.* High-Curie-temperature ferromagnetism in self-organized $Ge_{1-x}Mn_x$ nanocolumns. *Nature Mater.* **5**, 653-659 (2006).
46. Bougeard, D., Ahlers, S., Trampert, A., Sircar, N. & Abstreiter, G. Clustering in a precipitate-free GeMn magnetic semiconductor. *Phys. Rev. Lett.* **97**, 237202 (2006).
47. Winkler, E., Zysler, R. D., Vasquez Mansilla, M. & Fiorani, D. Surface anisotropy effects in NiO nanoparticles. *Phys. Rev. B* **72**, 132409 (2005).
48. Makarova, T. L. *et al.*, Magnetic carbon. *Nature* **413**, 716–718 (2001); Retraction, *Nature* **440**, 707 (2006).
49. Abraham, D. W., Frank, M. M. & Guha, S. Absence of magnetism in hafnium oxide films. *Appl. Phys. Lett.* **87**, 252502 (2005).
50. Mitome, M. *et al.* Nanoanalysis by a high-resolution energy filtering transmission electron microscope. *Microsc. Res. Tech.* **63**, 140-148 (2004).

**Acknowledgments**

This work was partially supported by the Grant-in-Aids for Scientific Research (Basic Research (B) and Priority Areas), the 21st COE program of the University of Tsukuba, and the 'Nanotechnology Support Project' of the Ministry of Education, Culture, Sports, Science and Technology (MEXT), Japan. We would like to thank N. Ozaki, T. Kumekawa, K. Kadowaki (Univ. Tsukuba), O. Eryu (Nagoya Institute of Technology) and T. Ohshima (Japan Atomic Energy Agency) for contributions and supports in the experiments. T. D. thanks A. Bonanni, F. Matsukura, H. Ohno, and M. Sawicki for valuable discussions. Correspondence and requests for materials should be addressed to S. K. and T. D. Supplementary Information accompanies this paper on www.nature.com/ naturematerials.

**Competing financial interests**

The authors declare that they have no competing financial interests.

**Figure captions**

**Figure 1 TEM and transmission electron diffraction pattern of a $Zn_{0.95}Cr_{0.05}Te$:I film. a,** Cross-sectional image of an I-doped $Zn_{0.95}Cr_{0.05}Te$ epilayer obtained by high-resolution transmission electron microscopy. The iodine concentration being $[I] = 2 \times 10^{18} cm^{-3}$ and the apparent Curie temperature $T_C^{(app)} = 300$ K. **b,c,** Electron diffraction images of the same epilayer: a region close to the interface with the ZnTe buffer layer (**b**); a region close to the surface (**c**).

**Figure 2 Comparison of Cr distributions in a series of $Zn_{0.95}Cr_{0.05}Te$ films.** Cross-sectional mapping images of the Cr $K_\alpha$ emission intensity (EDS mapping) for a series of $Zn_{1-x}Cr_xTe$ films with the same Cr composition $x = 0.05 \pm 0.005$ but differing in co-doping or stoichiometry. **a-c,** I-doped films with iodine concentrations of $[I] = 1 \times 10^{19} cm^{-3}$ (**a**); $[I] = 2 \times 10^{18} cm^{-3}$ (**b**) and $[I] < 1 \times 10^{17} cm^{-3}$ (**c**). **d,e,** Undoped films grown in Zn-rich (**d**) and Te-rich (**e**) conditions. **f,** N-doped film with a nitrogen concentration of $[N] = 4 \times 10^{20} cm^{-3}$.

**Figure 3 Characteristic temperatures of magnetic properties in the same series of $Zn_{0.95}Cr_{0.05}Te$ films.** The data of I-doped, undoped, and N-doped $Zn_{1-x}Cr_xTe$ films ($x = 0.05 \pm 0.005$), for which the Cr distributions are shown in Fig. 2, according to labels a-f. The apparent Curie temperature, $T_C^{(app)}$, deduced from the Arrott plot analysis is shown by red circles; the paramagnetic Curie-Weiss temperature, $\Theta_p$, is shown by green triangles and the blocking temperature, $T_B$, is shown by blue squares; $T_B$ represents the temperature at which the magnetization measured in 500 Oe after cooling in zero magnetic field attains a maximum. If the maximum is in the form of a plateau, its width is denoted by an error bar (see Supplementary Information, Fig. S2). The data for I-doped (left panel) and N-doped (right panel) films are plotted against the doping concentration of iodine or nitrogen, whereas the data for undoped films (middle panel) are plotted against the Te/Zn flux ratio.

**Figure 4 Different charge states of Cr in ZnTe and the energy for forming Cr pairs. a,** Schematic positions of the energy levels of Cr in two charge states: N acceptors and I donors in ZnTe, according to refs 35, 36, respectively. **b,** Result of *ab initio* calculations of the total energy change, $E_d$, by forming Cr-Cr pairs in the nearest-neighbour positions in the cation sublattice of ZnTe for Cr in various charge states.

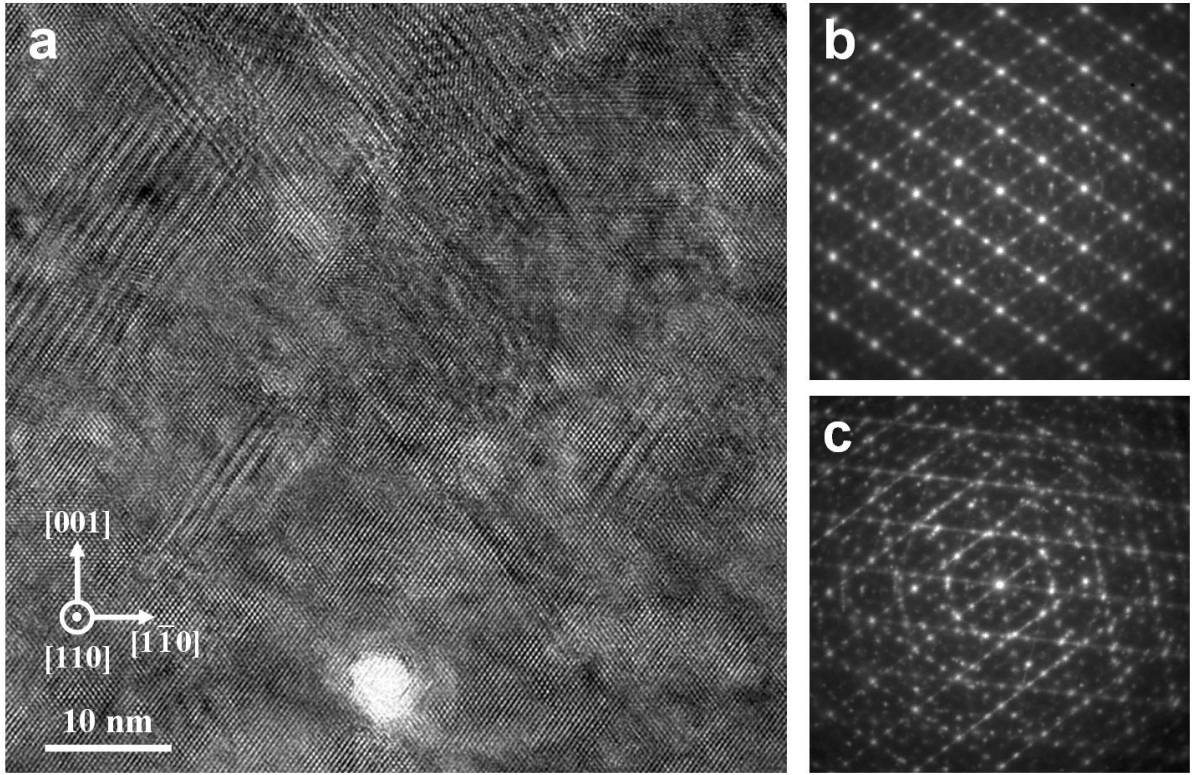

**Figure 1**

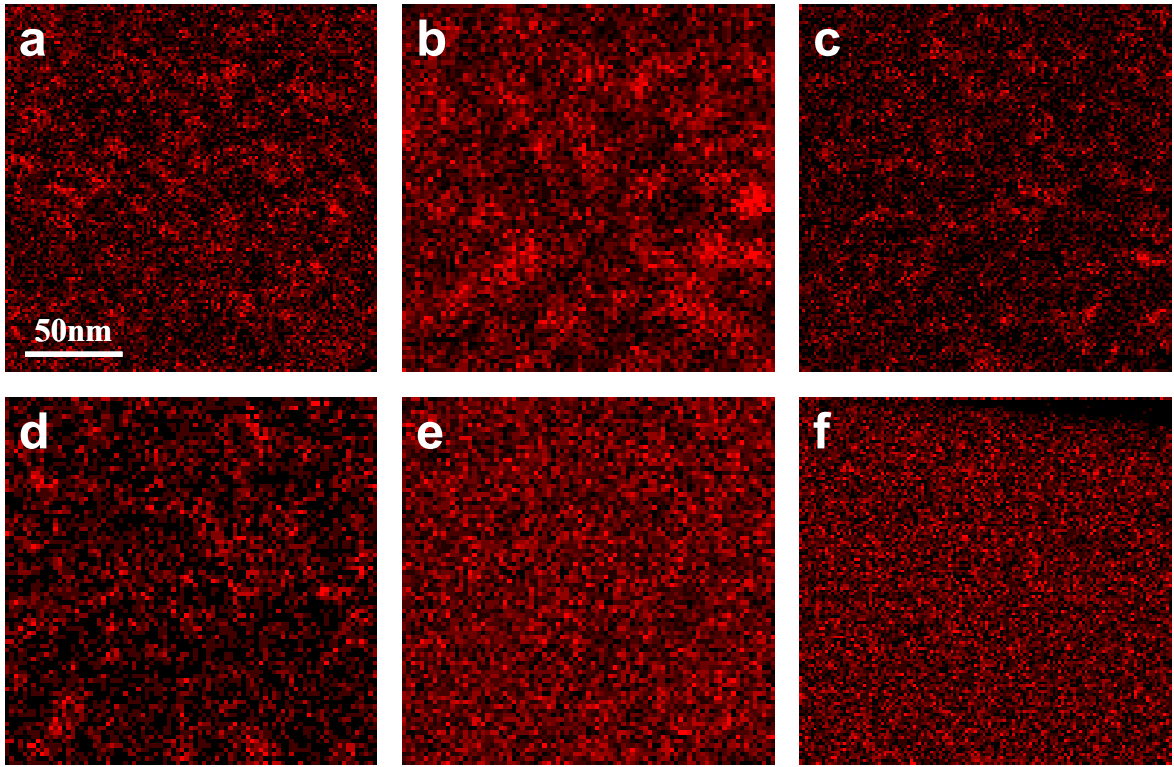

**Figure 2**

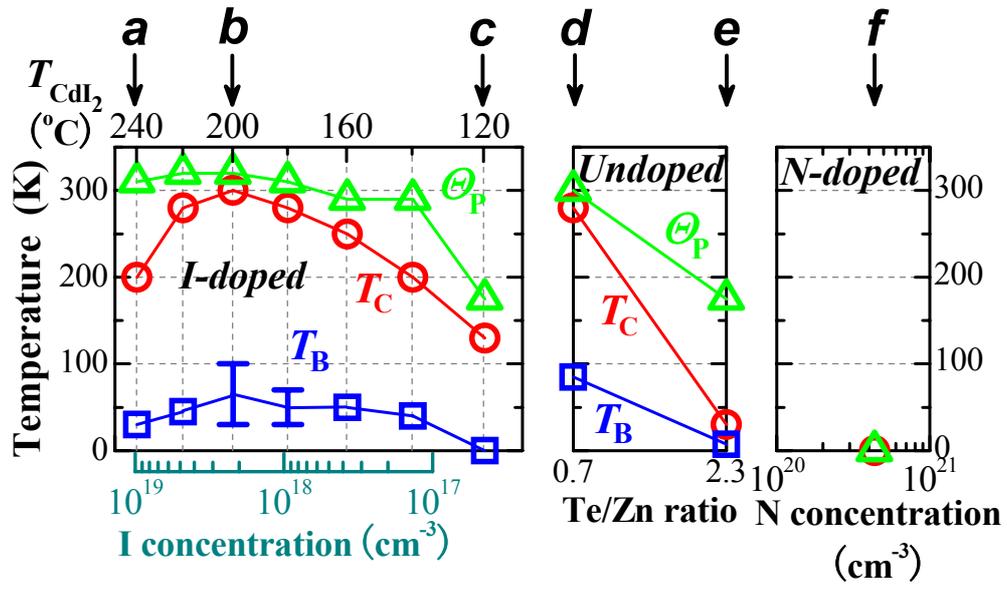

**Figure 3**

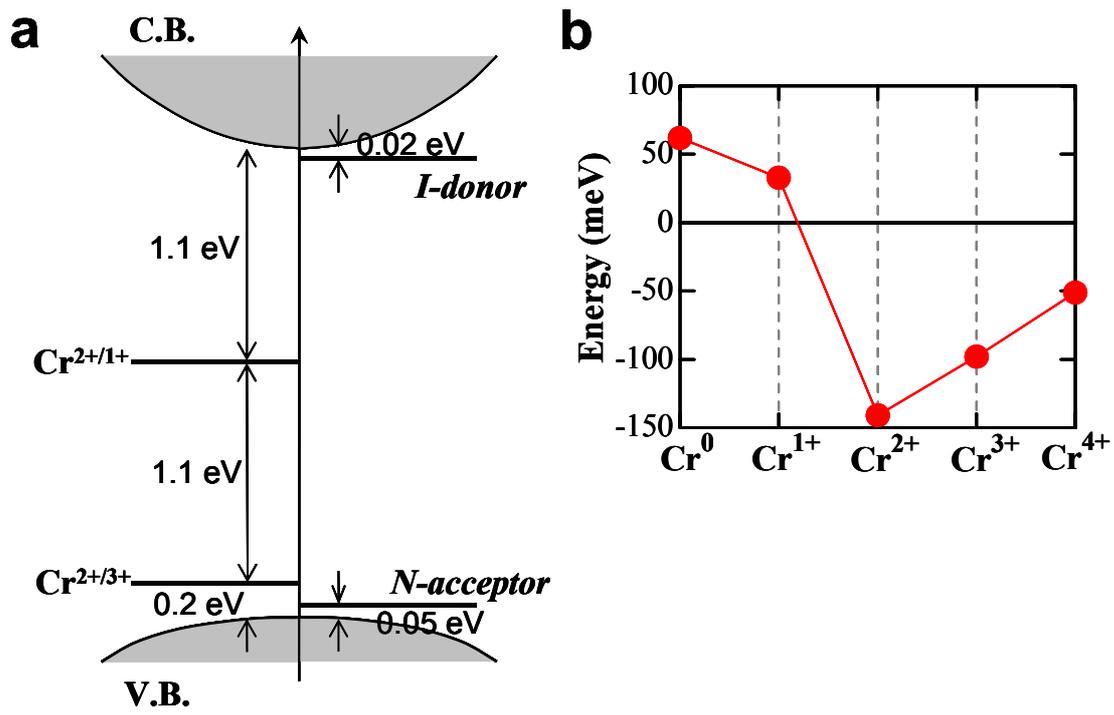

**Figure 4**

# Supplementary Information for
# "Origin and control of high-temperature ferromagnetism in semiconductors" – figures and legends

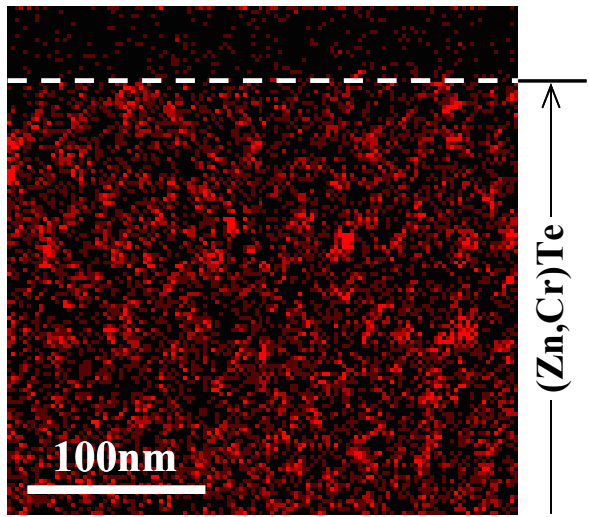

**Figure S1**

**Cr distribution probed in a thinner section of $Zn_{0.95}Cr_{0.05}$Te:I film.**

Cross-sectional mapping image of the Cr $K_\alpha$ emission intensity (EDS mapping) of the same $Zn_{0.95}Cr_{0.05}$Te:I film as in Fig. 2b, but probed in a thinner piece ($t \sim 50$ nm) prepared by Ar ion milling. Due to a reduced thickness over which the detected signal is averaged, we observe smaller sizes (10~20 nm) and a higher Cr composition ($x \sim 0.4$) of the Cr-rich regions (brighter spots). The nanograph reveals the development of uniformity in size and correlation in positions of the Cr-rich regions.

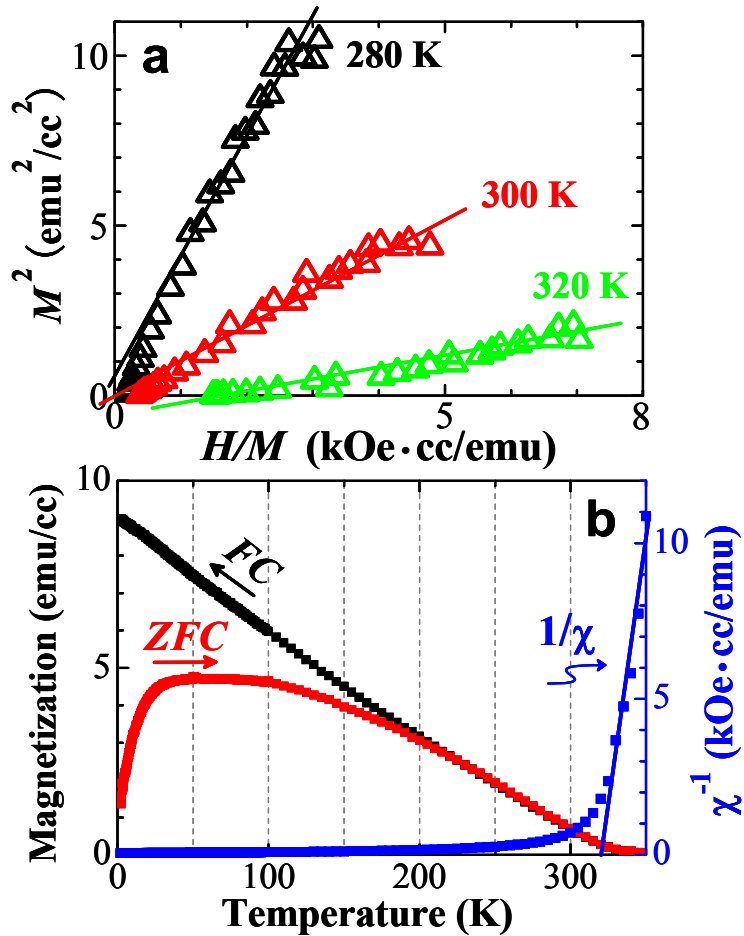

**Figure S2**

**A typical example of magnetization measurements on I-doped $Zn_{0.95}Cr_{0.05}Te$.**

**a,** Arrott plot analysis of magnetization vs. magnetic field for an I-doped $Zn_{0.95}Cr_{0.05}Te$ epilayer (sample b in Fig. 3), which leads to the apparent Curie temperature $T_C^{(app)}$ = 300 K. **b,** Temperature dependence of magnetization and inverse magnetic susceptibility for the same sample under the magnetic field of 500 G, which results in the paramagnetic Curie-Weiss temperature $\Theta_p$ = 320 K. Magnetization measurements taken after cooling in the absence of an external magnetic field (ZFC process) exhibit a flat maximum in the temperature range of 30~100 K, which is represented by the blue bar in the plot featuring the blocking temperature $T_B$ in Fig. 3.

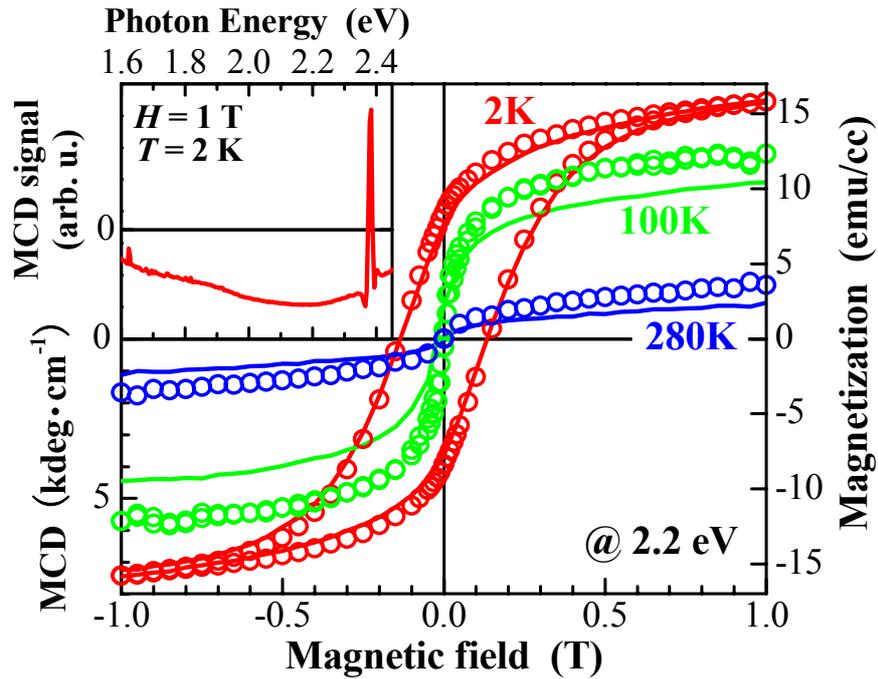

**Figure S3**

**Similar magnetic-field dependences between MCD and magnetization in I-doped $Zn_{0.93}Cr_{0.07}Te$.**

The magnetic-field dependence of MCD signal intensities (represented by lines) of an I-doped $Zn_{0.93}Cr_{0.07}Te$ epilayer ($T_C$ = 280 K, $T_{CdI2}$ = 200°C). The measurement was performed in the transmission mode under magnetic fields perpendicular to the layer. The magnetization measured using SQUID (represented by circles) showed similar field dependences to those of MCD. The inset shows the MCD spectra at $H$ = 1 T and $T$ = 2 K.

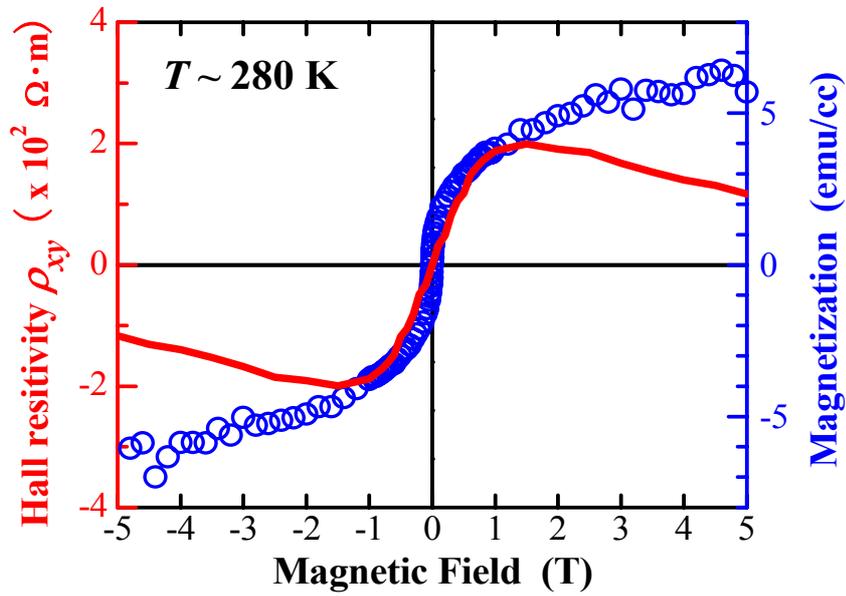

**Figure S4**

**Hall resistivity $\rho_{xy}$ and magnetization of I-doped $Zn_{0.95}Cr_{0.05}Te$.**

Hall resistivity $\rho_{xy}$ (represented by red curve) and magnetization (blue circles) as a function of magnetic field for an I-doped $Zn_{0.95}Cr_{0.05}Te$ epilayer (sample b in Fig. 3). The Hall measurement was performed at $T$ = 275 K. (The measurement at lower temperatures was hindered by a significant increase of resistivity with decreasing temperature.) Below 1 T, the anomalous Hall effect dominates Hall conductivity $\rho_{xy}$, which shows a similar field dependence to those of magnetization. At higher fields, a contribution of the ordinary Hall effect becomes clear.